\documentclass{ase}
\usepackage{graphicx}
\usepackage{amssymb}
\usepackage{mathtools} 
\usepackage{caption}
\usepackage{algorithm}
\usepackage{algpseudocode}
\usepackage{float}

\begin{document}

\begin{center}
{\huge Hoodsquare: Modeling and Recommending Neighborhoods in Location-based Social Networks}


\begin{multicols}{4} 
{Amy X. Zhang}{\\}{Massachusetts Institute of Technology\\\texttt{axz@mit.edu} \\ \columnbreak}
{Anastasios Noulas}{\\}{University of Cambridge\\\texttt{an346@cam.ac.uk} \\ \columnbreak}
{Salvatore Scellato}{\\}{University of Cambridge\\\texttt{ss824@cam.ac.uk}\\ \columnbreak}
{Cecilia Mascolo}{\\}{University of Cambridge\\\texttt{cm542@cam.ac.uk}\\}
\end{multicols}
\end{center}

\begin{multicols*}{2}
\begin{abstract}
Information garnered from activity on location-based social networks can be harnessed to characterize urban spaces and organize them into neighborhoods. In this work, we adopt a data-driven approach to the identification and modeling of urban neighborhoods using location-based social networks. We represent geographic points in the city using spatio-temporal information about Foursquare user
check-ins and semantic information about places, with the goal of developing features to input into a novel neighborhood detection algorithm. The algorithm first employs a similarity metric that assesses the homogeneity of a geographic area, and then with a simple mechanism of geographic navigation, it detects the boundaries of a city's neighborhoods. The models and algorithms devised are subsequently integrated into a publicly available, map-based tool named Hoodsquare that allows users to explore activities and neighborhoods in cities around the world. 

Finally, we evaluate Hoodsquare in the context of a recommendation application where user profiles are matched to urban neighborhoods. By comparing with a number of baselines, we demonstrate how Hoodsquare can be used to accurately predict the home neighborhood of Twitter users. We also show that we are able to suggest neighborhoods geographically constrained in size, a desirable property in mobile recommendation scenarios for which geographical precision is key.
\end{abstract}

\section{Introduction}
\label{sec:intro}
\begin{table*}
\centering
\begin{tabular} {| l | r | r | r | }
\hline
City & Venues &  Users & Check-ins\\
\hline
London & $14,137$ & $8,132$ & $341,829$ \\
\hline
New York & $39,058$ & $17,833$ & $923,259$ \\
\hline
San Francisco & $21,533$ & $20,971$ &  $235,898$ \\
\hline
\end{tabular}
\caption{Volume for each Foursquare data entity and city used in our study.}
\label{tabledata}
\end{table*}
Neighborhoods have been in existence for as long as there have been areas for people to congregate in. They play an important role in segmenting activities, industries, and people within urban spaces.
The knowledge of the neighborhood location of a user is also pivotal to the development of services such as recommendation systems able to advise on visits to locations such as restaurants, museums, and shops. 
As a result of the expansion of mobile applications, these systems have become an important way for users to discover and explore local areas.
At the same time, mobile applications have enabled the collection of data about users' behavior and mobility with an unprecedented spatial, semantic, and temporal granularity. However, research into algorithms and models that are able to integrate these elements is still lacking.
\begin{figure*}
\centering
{\includegraphics[width=1.0\columnwidth]{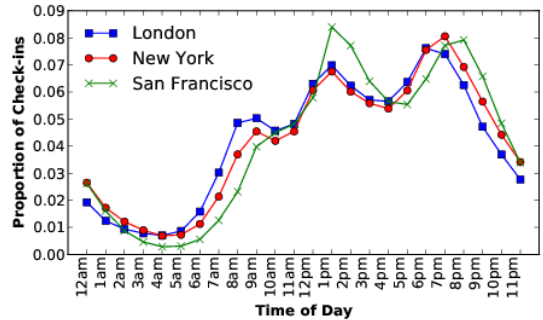}}
\caption{Hourly check-in volume in the three cities, normalized by the total number of check-ins for that day, averaged over all days.}
\label{fig:hourlyplot}
\end{figure*}

In this paper, we mine a dataset crowdsourced from the most popular online location-based service, Foursquare and redefine the notion of a neighborhood within a dynamic spatio-temporal context by means of user ``check-ins''. By exploiting multi-dimensional representations of user check-in activity across the geographic space, we devise an algorithm for extracting neighborhood boundaries in cities. Subsequently, we integrate the output into a map-based tool tailored for urban exploration and showcase its usability in a mobile recommendation scenario.
More in detail, our contributions are the following:\\
\\
\textbf{Urban Activity Feature Mining.} By performing analysis on the spatio-temporal dynamics of Foursquare user check-ins in three metropolitan areas (New York, London, and San Francisco), we identify a range of data mining features that can be used to characterize urban user activity. We define three broad classes of features: those based on Foursquare venue types (Shops, Nightlife, etc.), on the spatial congregation of \textit{locals} and \textit{tourists} in the city, and on temporal information about user check-ins. In total, more than three hundred features are used to offer a multi-dimensional representation of the urban territory and build the input for a neighborhood detection algorithm.\\
\\
\textbf{Neighborhood Detection.} The feature vectors employed to represent granular geographic points in the city are then used to detect urban neighborhoods. The mining process is carried out in three well-defined steps that lead to the extraction of geographic boundaries that encapsulate homogeneous pockets of land in the city. First, the relative importance of a feature at an area is decided by employing the OPTICS algorithm~\cite{Ankerst_1999}. OPTICS operates on spatial data and returns clusters that correspond to areas of high relative density for a given feature. Next, we devise a similarity metric of homogeneity across the city's geographic points and finally, we integrate the metric into a neighborhood extraction algorithm that, 
given as input the set of points in a city, returns a set of convex polygons, each corresponding to an urban neighborhood. The algorithm infers neighborhoods by employing a simple mechanism of geographic navigation. It allows for the generation of controllable neighborhood sizes, while at the same time is able to cope with any density heterogeneities that may exist in the territory of a city.
\newline  
\\
\textbf{Urban Exploration and Neighborhood Recommendation.} 
The results of the features and algorithms we develop are presented in an online, map-based tool called Hoodsquare\footnote{\url{www.hoodsquare.org}}. Users can navigate through a set of cities around the world, exploring various urban activities as well as discovering neighborhoods and what activities are prominent in each of them. 
We finally evaluate the techniques developed in a mobile recommendation scenario that aims to predict a user's home neighborhood (the neighborhood he or she visits most) based on textual Twitter profile data. We compare our prediction results with other neighborhood detection baselines defined by a real estate service, U.S. census tract data, and the Livehoods project presented in~\cite{Cranshaw_2012}. We show that the Hoodsquare algorithm can be used to recommend geographic areas that are small in size and that maintain a balanced trade-off between prediction accuracy and geographic precision. 

Hoodsquare has many potential applications. Beyond greater societal awareness of our cities, urban planning as a discipline can take advantage of this framework to better understand the growth and development of cities, including phenomena such as urban gentrification~\cite{gentrification}.
In addition, the identification of salient characteristics and accurate location of neighborhoods is important to venue and location recommendation applications.
The spirit of our work is in line with the recently launched application by the online rental service, Airbnb~\cite{airbnb}, which enables web users to browse interesting neighborhoods across cities when searching for accommodation away from home.

\section{Urban Feature Mining}
\label{sec:analysis}
\begin{figure*}[t]	
\centering
\subfigure[NYC Morning]
{
\label{combinedapr}
\includegraphics[width=0.65\columnwidth] {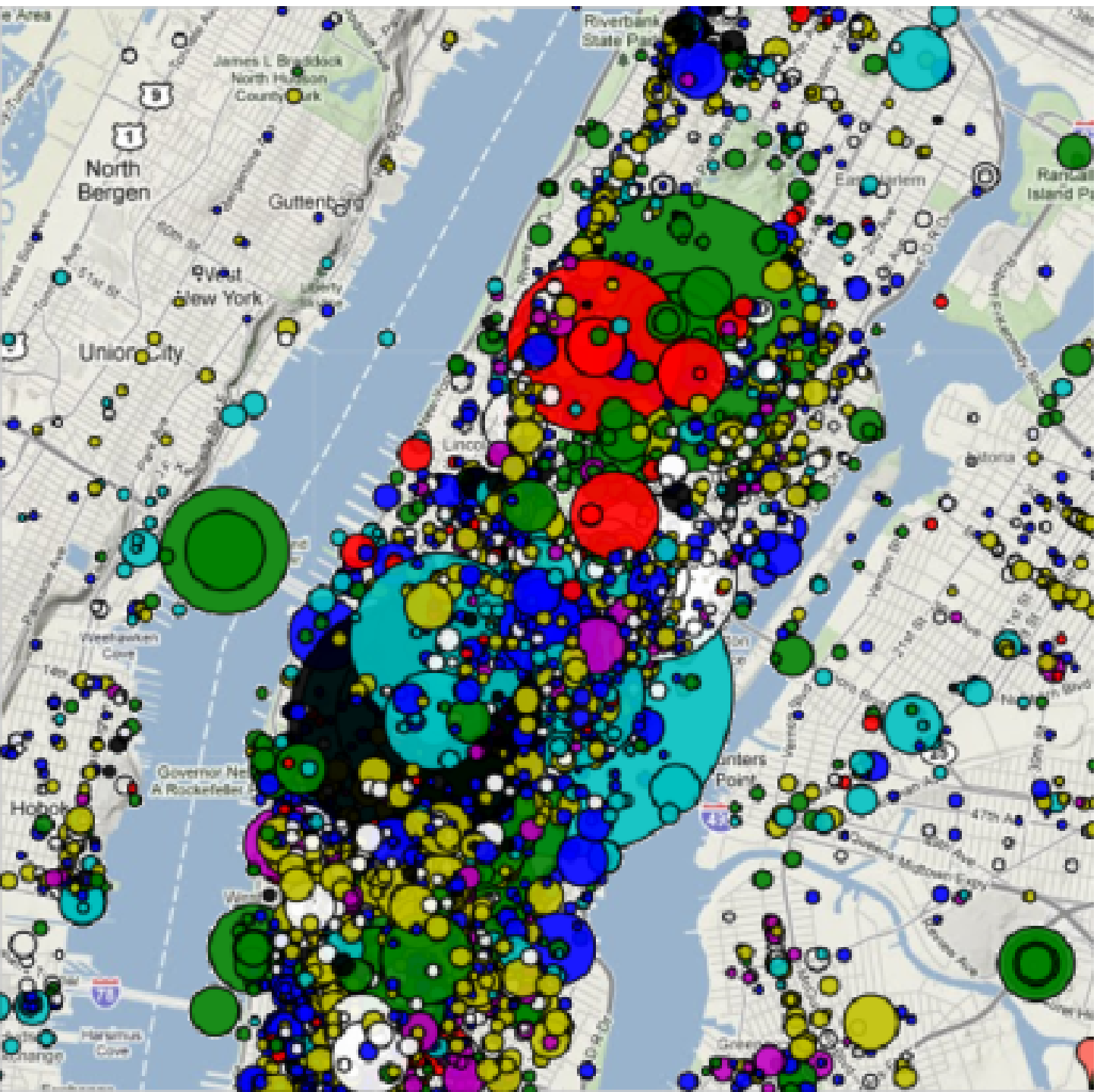}
}
\subfigure[NYC Night]
{
\label{combinedaccuracy}
\includegraphics[width=0.65\columnwidth] {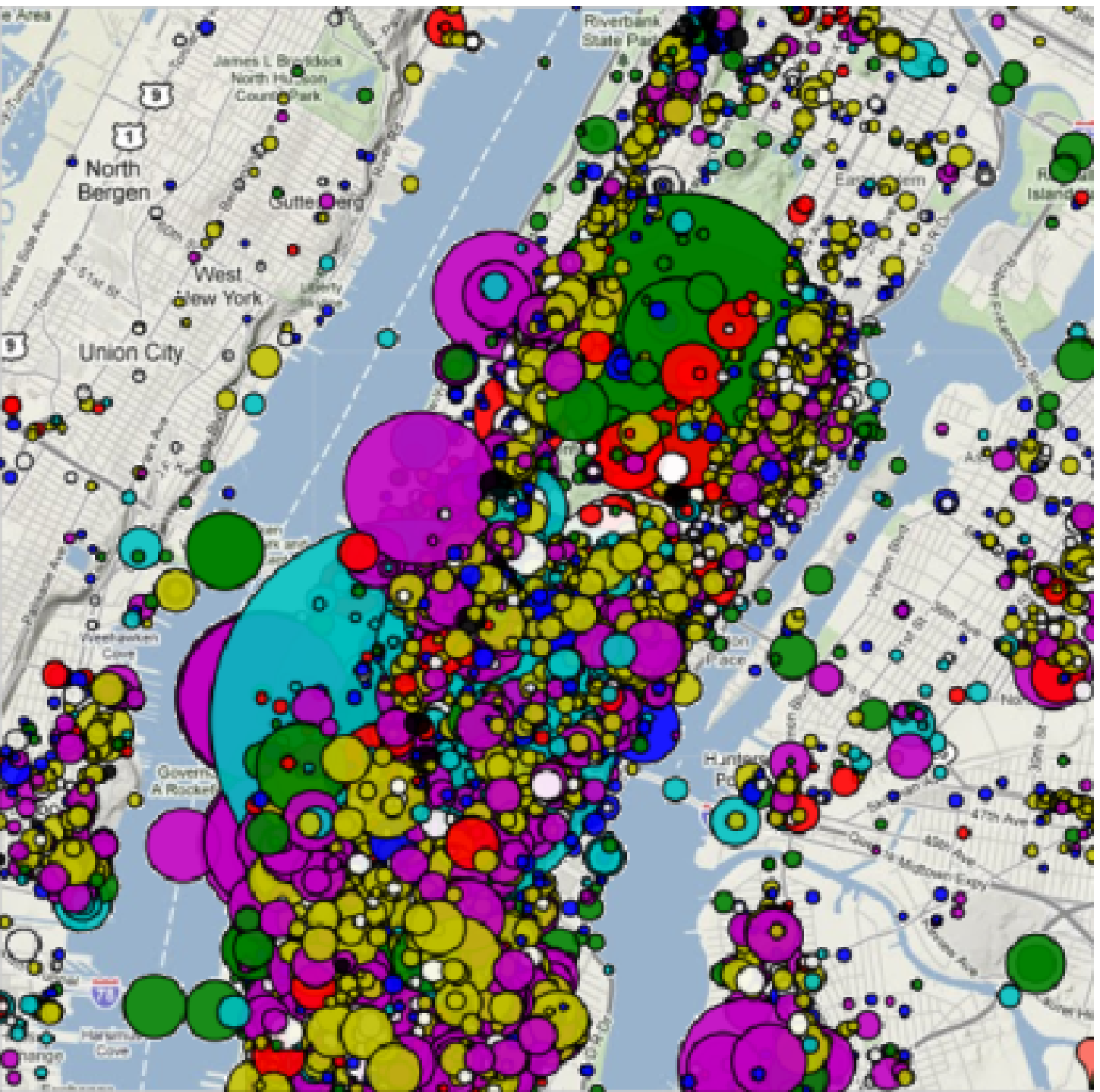}
}
\caption{Urban activity comparison between morning and night hours in Manhattan. Each circle corresponds to a Foursquare venue. Colors are representative of different venue types: Arts (red), Education (black), Shops (white), Food (Yellow), Parks (green), Travel (cyan), Nightlife (magenta), Work (blue). The radius of a circle is proportional to the popularity of each venue.}
\label{fig:NYtime}
\end{figure*}

The dataset we employ for this work spans approximately five months, dating from May 27th, 2010 to November 2nd, 2010 and includes check-ins from Foursquare accounts forwarded to public Twitter profiles.
This accounts for approximately $20\%$ of total Foursquare check-in traffic for the stated period, and information on the general properties of the dataset can be found in~\cite{noulas2011empirical}.
In Table~\ref{tabledata} we present a summary of relevant information for the three cities (New York, London, and San Francisco) we focus our analysis on,
although in the beta version of Hoodsquare accessible online, we integrate 11 more cities around the world. To associate check-ins with each city, we retrieve the corresponding bounding-box information from the Yahoo!\ PlaceFinder web service\footnote{\url{developer.yahoo.com/geo/placefinder/}}. For every place, we acquire its latitude and longitude coordinates as well as the name and place category, such as Bar or Restaurant. For each user, we also have his or her Twitter username, profile information, and a set of check-in locations along with the time each check-in took place.
\subsection{Preliminary Analysis}
We now perform an analysis to highlight the important characteristics of the check-in dataset with respect to the modeling of urban spaces.
Considering approximately $300$ venue categories that exist in Foursquare, we note that the presence of top place categories in each city is different. For instance in London, the category with the greatest number of places is Pub with $6.6$\%, while in New York, it is Corporate/Office with $6.2$\%, and in San Francisco, $4.3$\% of venue types are home residences. Taking into account the volume of check-ins, however, paints a different picture of where people actually check in. In San Francisco, $48.2$\% of check-ins are at airports, while in New York and London, $22.2$\% and $33$\% of check-ins, respectively, are at train stations. 
This shows that Foursquare check-ins may be biased towards travel places as people may check-in more when they are traveling, but also highlights that user behavior may be driven by the cultural idiosyncrasies of the respective home cities.

Focusing on the spatio-temporal evolution of Foursquare activity, when analyzing user check-ins across the 24-hour daily cycle, we find that check-ins spike around the time normally associated with mealtime, as can be seen in Figure~\ref{fig:hourlyplot}. Small variations are noticeable across the three different cities, but in qualitative terms the volume of check-ins evolves similarly. In Figure~\ref{fig:NYtime}, we depict two snapshots of check-in activity near Manhattan over morning and nighttime, respectively. Different categories of places (as highlighted by the different colors) become more dominant in terms of check-in frequency at different temporal intervals, as well as across the geographic space. This provides a good indication of how the spatio-temporal dynamics of Foursquare user check-ins could be employed by algorithms to characterize urban landscapes, one of the principal aims in the present work. 

\begin{figure*}
\centering
{\includegraphics[width=1.0\columnwidth]{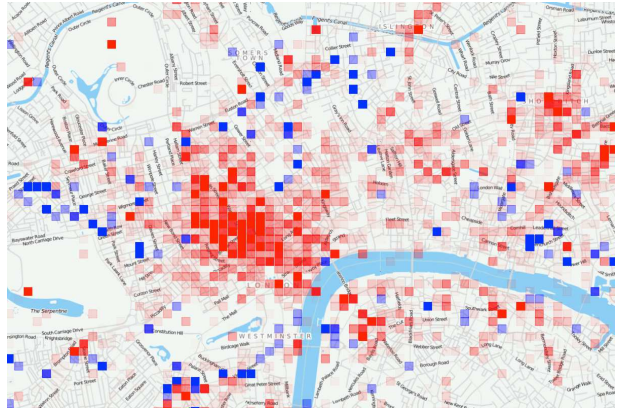}}
\caption{Visualization of check-ins by tourists and locals in London, with stronger blue indicating more tourist check-ins and stronger red more local check-ins.}
\label{fig:LDNTL}
\end{figure*}

Further, an interesting attribute useful to characterize neighborhoods is touristic activity, since the socio-economic identity and cultural fingerprint of an urban area may be influenced by the presence of tourists. Thanks to the global coverage of Foursquare, we are able to observe the geographic spread of check-ins for a particular user and tag him as a \textit{tourist} or \textit{local} for a city. We tag a user as a local in a city if $50$\% or more of his or her check-ins during the time period of data collection are within the boundaries of the city and as a tourist otherwise. By plotting the check-ins of tourists and locals in London in Figure~\ref{fig:LDNTL}, we observe that some areas receive more tourist visitors while others tend to attract local users. Next, we demonstrate how the signal, sourced from the check-in data analyzed above, can be mined to form machine learning features that are then used for neighborhood detection.
\subsection{Urban Activity Features}
\begin{figure*}
\centering
{\includegraphics[width=1.3\columnwidth]{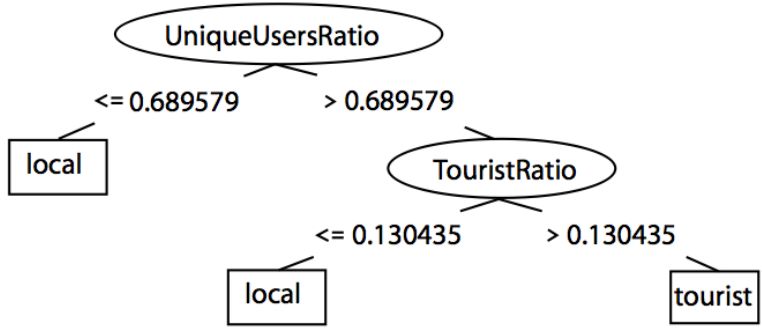}}
\caption{Decision Tree that has emerged after training the C4.5 algorithm with a set of training instances that correspond to Local and Tourist places in London.}
\label{fig:tree}
\end{figure*}

Driven by the observations described in the previous section we build a set of data mining features to model various aspects of urban activity revealed by Foursquare user check-ins. 
The local homogeneity of these characteristics in addition to others such as demographic and economic makeup have been found to define neighborhoods, as reflected in a large body of research on geographical groupings of similar places, activity, and people \cite{Hotelling_1929,  Tiebout_1956}. We note that while we attempt to be exhaustive in this task and mine a large spectrum of relevant features, we acknowledge that alternative formulations or data types could also be considered. Indeed, the algorithmic models to be presented in Section~\ref{sec:model} are easily generalizable to new features.
\newline
\\
\textbf{Foursquare Venue Types.}
Given the set of Foursquare venues in a city, we associate each with a specific \emph{Place Type}. The data we have collected from Foursquare already provides semantic annotations of place types for each venue. This task has been crowdsourced by Foursquare users, while the taxonomy of venue types and the hierarchical tree of categories has been provided by Foursquare. We use this tree of category organization, a complete directory of which can be found on the Foursquare website~\cite{fourcats}, to represent venues with different levels of abstraction. For instance, we could have a Starbucks venue and represent it with categories such as Coffee Shop and Food, since a coffee shop is also a type of food venue. Overall, the almost $300$ venue types available in Foursquare provide a source for representing a wide spectrum of activities in the city, and as we will see in the coming sections, they constitute a basic ingredient to characterizing neighborhoods.
\newline
\\
\textbf{Temporal Features.}
Next, we represent each place with a \emph{Time} feature that signifies the busiest time of the day for the place. From the check-ins collected in a city, we create for each place $p$, the time series, $T24_{p}(h)$ for $h = 0,...,23$ by counting all the checkins for every hour in a 24-hour day, restricted to places that have at least 6 check-ins.
Then we apply on $T24_{p}(h)$ a Gaussian smoothing function. 
From the 24-hour time series of smoothed values for each place, we select the maximum position, giving us the hour of the day 0--23 that the place is busiest. We can now define a set of \emph{Time} features that represent different periods of a day. We use Figure~\ref{fig:hourlyplot}, which shows increased volume during mealtimes, to guide our segmentation of time, as there is no standard. The following tree shows the set of temporal features we have employed:
\begin{quote}
{\small
\begin{tabbing}
Daytime: \=6AM - 6PM \\
\>	Morning: \=6AM - 11AM \\
	\> \>	 Breakfast: 7AM - 10AM  \\
\>	 Midday: 10AM - 2PM \\
	\>\>	 Lunch: 11AM - 1PM \\
\>	 Afternoon: 1PM - 5PM \\
 Nighttime: 6PM - 6AM		\\
	\> Dinner: 6PM - 9PM \\
         \>  Late evening: 8PM - 2AM\\
	\>\>	Midnight: 10PM - 2AM\\
\>	Early morning: 2AM - 6AM
\end{tabbing}}
\end{quote}
We tag a place with one of the temporal periods, if the hour with the greatest volume of check-ins lies between its range.
\newline
\newline
\textbf{Local and Tourist Venue Classification.}
In order to define a feature that identifies a Foursquare venue as local or touristic, we devise a supervised learning classification method.
In this context, we have manually labeled 33 places in each city as touristic, such as museums, attractions, and monuments that are known to attract touristic crowds - for instance, Times Square in New York and Buckingham Palace in London.
We then tag places with the Foursquare category of Home as local places.
In addition to labels, we have also derived two features which have been considered relevant for the classification of places as local or touristic. 
First, as mentioned in Section~\ref{sec:analysis}, we have categorized a user that checks in at a city as a tourist, if more than half of their check-ins are outside the corresponding urban boundary. We subsequently measure the \emph{TouristRatio} of a place, that is, the number of tourists that check in to a place normalized with respect to the total number of users observed at that place. 
Second, touristic places generally have more one-time visitors, while local places are more likely to be revisited by their users. We thus calculate the \emph{UniqueUsersRatio} metric for each place by counting the proportion of unique users that check in to a place. 

Having set labels and the two features, we use supervised learning classifiers that can learn a function to classify a Foursquare venue as local or tourist. To do so we train a separate C4.5 decision tree classifier~\cite{c45tree} for touristic and local places of each city by exploiting the WEKA~\cite{WEKA} machine learning framework to classify unlabeled Foursquare venues as touristic or local. 
To give an indication of the general performance achieved, using a balanced dataset to train on $33$ tourist and $33$ local places where a random selection would provide $50$\% accuracy, our decision tree correctly classified instances $96.92$\% of the time using a $10$-fold cross validation technique. 
The decision tree used to classify venues in the city of London is depicted in Figure~\ref{fig:tree}. As expected, the classifier cannot guarantee that all real touristic spots will be detected, yet it provides a low cost solution, compared to manual labeling, that helps with the detection of places in the city that tend to attract tourists.

\section{Neighborhood Inference Model}
\label{sec:model}
\begin{figure*}[t]	
\centering
\subfigure[400m]
{
\includegraphics[width=0.275\columnwidth] {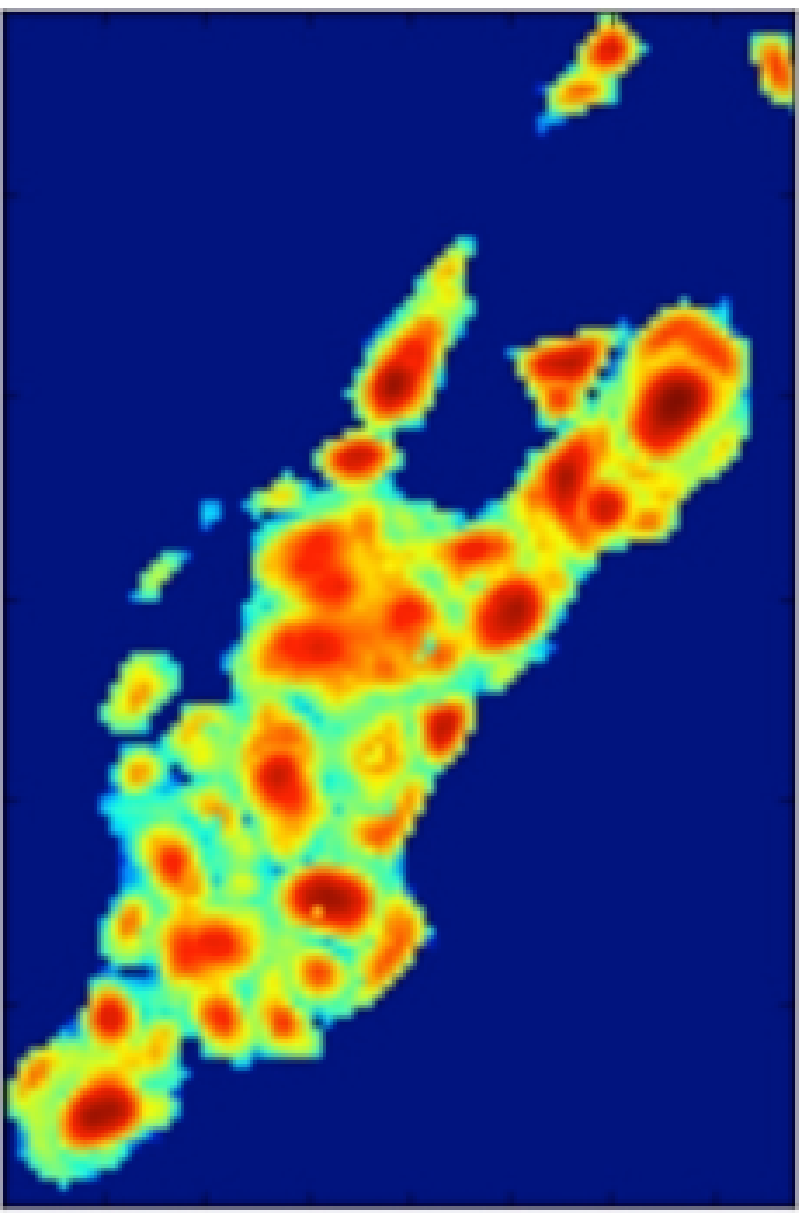}
}
\centering
\subfigure[800m]
{
\includegraphics[width=0.275\columnwidth] {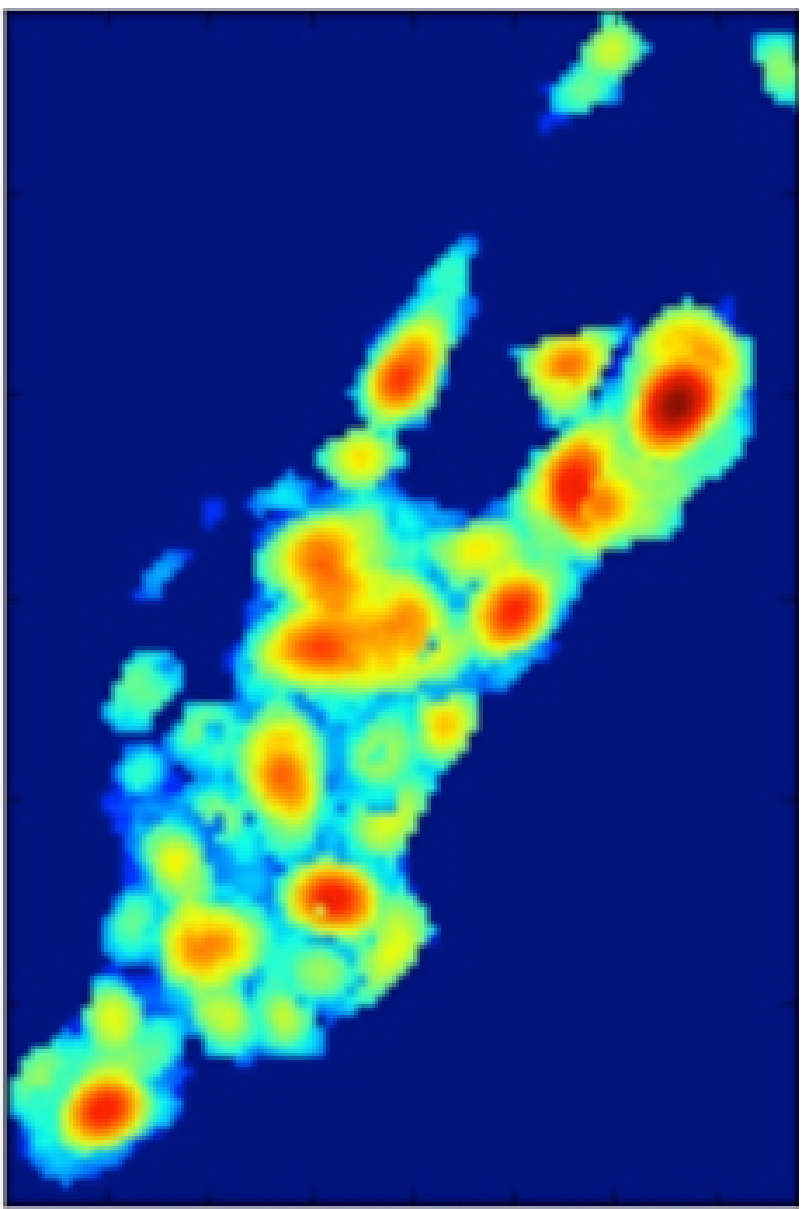}
}
\subfigure[1200m]
{
\includegraphics[width=0.275\columnwidth] {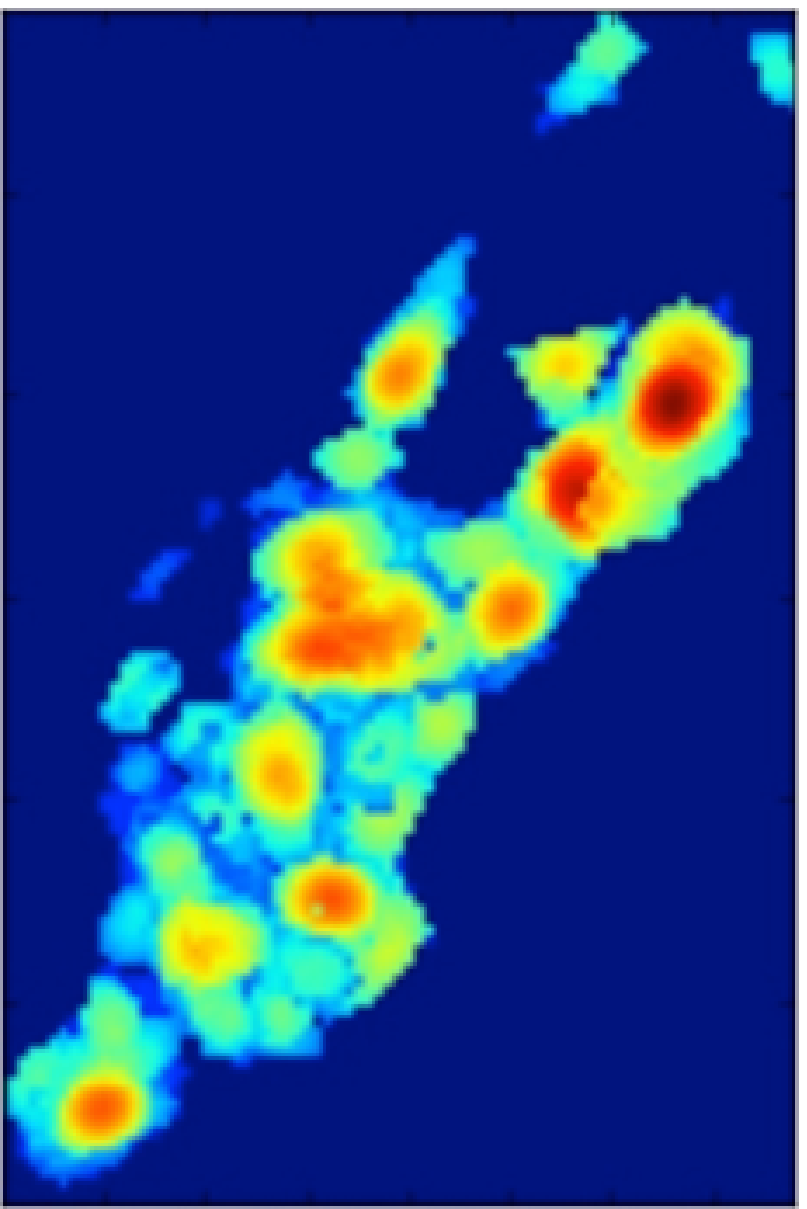}
}
\caption{Heatmaps for New York of \texttt{H\_Index} using radii of $400$, $800$, and $1200$ meters.}
\label{fig:Rad}
\end{figure*}

Given the features explored in the previous section we now present a model to detect and define neighborhoods or, put otherwise, localized areas of homogeneous characteristics. 
As a preprocessing step we segment the city into geographic cells by superimposing a grid over it.
$Grid$ cells have a width of $100$ meters on the latitudinal and longitudinal axes.
In order to gain an understanding of a cell's ambiance or character, we next devise a method to characterize the area represented by each cell,
exploiting the features we introduced previously. Ultimately, every cell $p \in Grid$ is formulated as a vector $\mathbf{v_p}$, with each dimension representing a feature. 
Then, the neighborhood detection methodology we developed can be explained using the following steps:
\begin{itemize}
\item{\textbf{Activity hotspot detection} (see Paragraph~\ref{modelStep1}): a spatial clustering algorithm is used to infer if each geographic point $p$ in the city is a \textit{hotspot}, or is locally dense, with respect to a particular feature (i.e., touristic or nightlife hotspot). While hotspots can be seen also as a separate component of the Hoodsquare website, they constitute a vital step to the detection of neighborhoods.}
\item{\textbf{Measuring area homogeneity} (see Paragraph~\ref{modelStep2}): a similarity metric is devised to assess if a given point $p$ is ``like'' its neighboring points in the city. The more hotspots two nearby geographic points share, the more similar they will be.} 
\item{\textbf{Neighborhood detection} (see Paragraph~\ref{modelStep3}): Finally, given as input the set of points in a city characterized in the previous steps, a boundary extraction algorithm returns 
a set of convex polygon groupings that represent our neighborhoods.} 
\end{itemize}
Our approach is explained in detail in the following paragraphs. 
\subsection{Identifying Activity Hotspots Through Spatial Clustering}
\label{modelStep1}
To compute a vector $\mathbf{v_p}$ for each cell $p \in Grid$, we first consider,
independently for each feature, all places in the city. For instance, if we take into account the feature \textit{Italian Restaurant}, we gather all Italian restaurants in the city represented via their geographic coordinates on the $2$-dimensional plane. 
Using as input these sets of places, we apply a density-based clustering algorithm to find hotspots for the feature, where hotspots are locally dense areas in the city with respect to the corresponding urban activity. 
We use the OPTICS algorithm~\cite{Ankerst_1999} to define the set of convex polygons that encapsulate areas of high relative density for the feature in question. OPTICS is chosen as it is developed specifically to deal with spatial data, does not require setting the number of clusters as an explicit parameter, and perhaps most importantly, can cope with clusters of varying densities which is 
a common case in an urban environment. 
\begin{figure*}[t]
\centering
\subfigure[Density Cos]
{
\includegraphics[width=0.275\columnwidth] {NewYork_Cos2_400.eps}
}
\centering
\subfigure[Binary Cos]
{
\includegraphics[width=0.275\columnwidth] {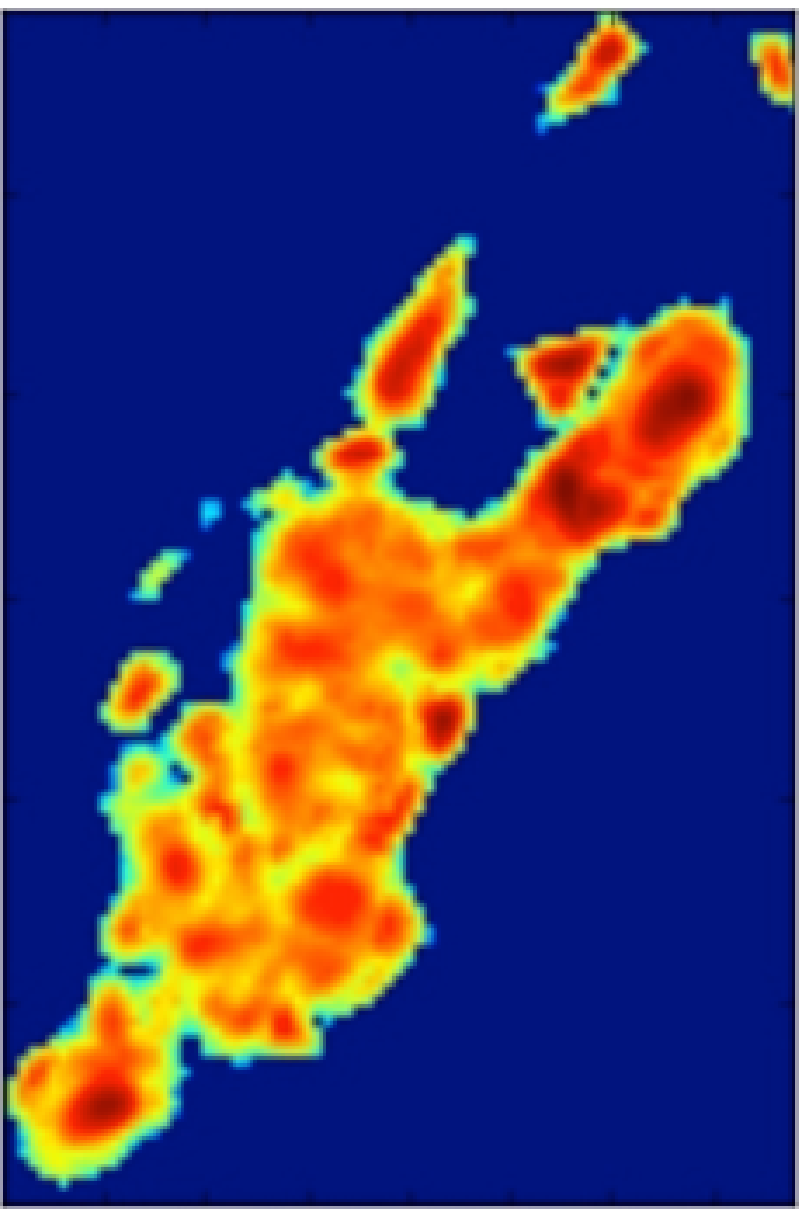}
}
\subfigure[Jaccard]
{
\includegraphics[width=0.275\columnwidth] {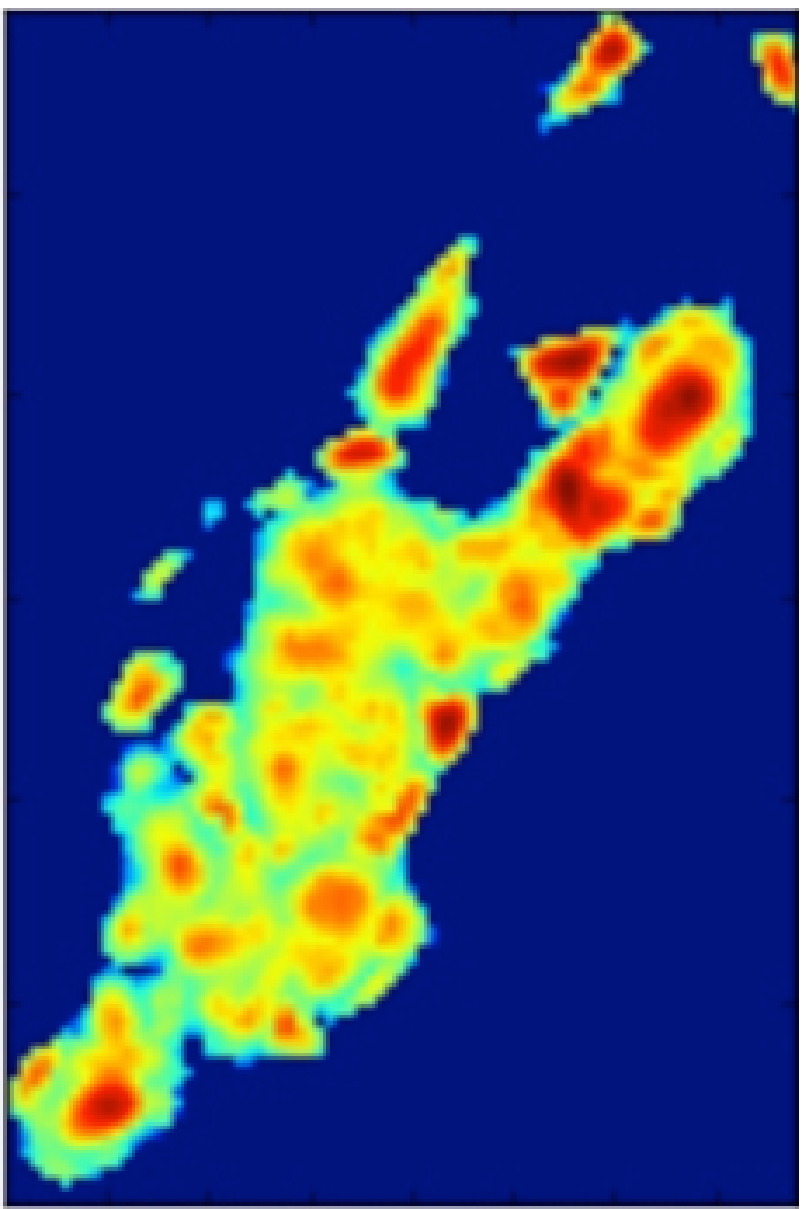}
}
\subfigure[Intersect]
{
\includegraphics[width=0.275\columnwidth] {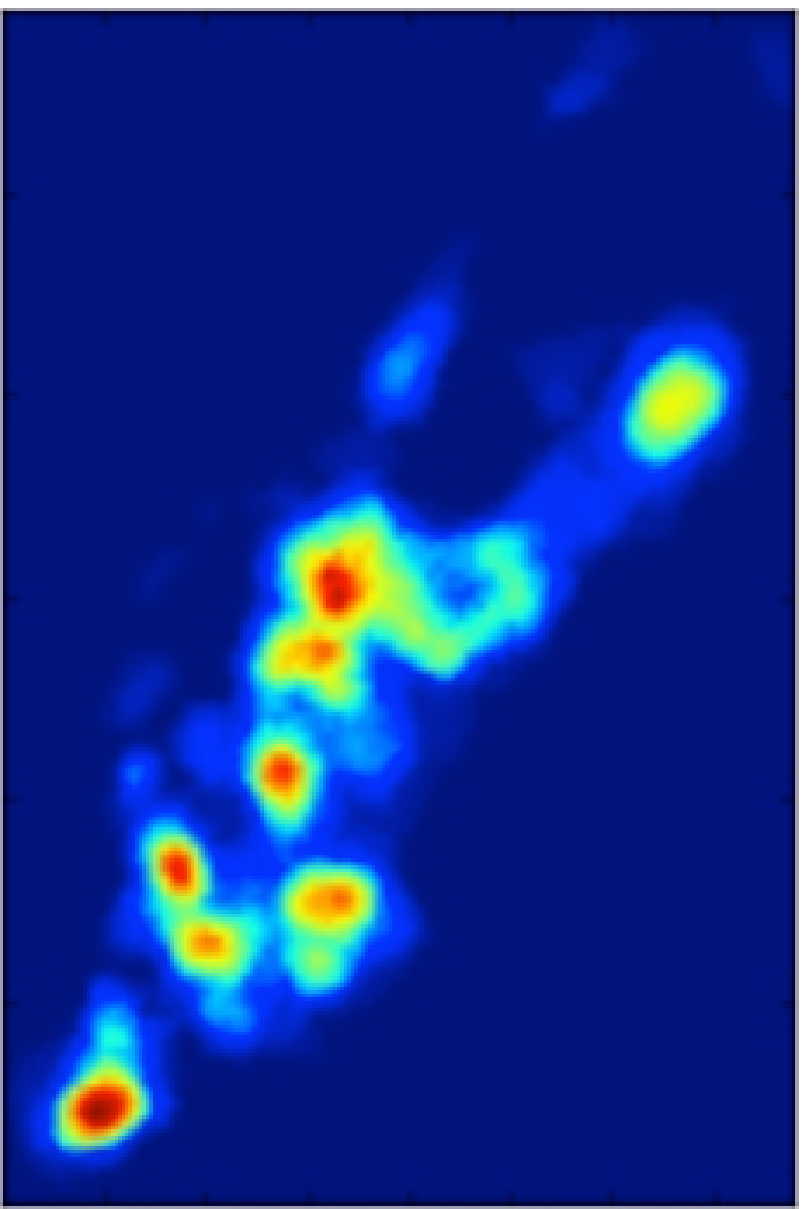}
}
\caption{Heatmaps for New York of \texttt{H\_Index} using four different similarity measures and a radius of 400 meters.}
\label{fig:Sim}	
\end{figure*}
As OPTICS returns a reachability plot instead of actual cluster memberships, we implement a modified version of an automatic clustering algorithm developed by Sander et al. \cite{Sander_2003}, which creates cluster trees containing only the significant clusters from a reachability plot. By then gathering the leaves off of this cluster tree and using the QuickHull algorithm to find the convex hull of each cluster, we obtain polygons that encapsulate the hotspots for each feature.
Given the full set of geographic polygons for our features we can now initialize the vector values $\mathbf{v_p}$ for each cell $p \in Grid$, with each dimension of $\mathbf{v_p}$ representing a different feature.
For each dimension of $\mathbf{v_p}$  we first check if $p$ is geographically covered by any of the polygons representing a hotspot for that feature. If it 
is not, then the value is set to $0.0$. Otherwise, we set the value of each feature according to the relative density of that feature to the polygon's area. That is, if we have $5$ venues characterized as touristic within the hotspot polygon, and its area is $0.25$ km$^2$, the value of the \emph{Tourist} feature will be $5.0/0.25 = 20.0$. In total, we have $310$ dimensions in $\mathbf{v_p}$ made up of $298$ \emph{Place Type} features, $10$ \emph{Time} features, and $2$ \emph{Local/Tourist} features.
\subsection{Geographic Cell Homogeneity}
\label{modelStep2}

An overall consensus among urban studies and public policy researchers
defines a ``neighborhood'' as \emph{a contiguous geographic area within a larger city, limited in size, and somewhat homogeneous in its characteristics} \cite{Chaskin_1997,Weiss_2007}.
Thus, given a collection of cells with each cell represented by the various features for which it is a hotspot, the next step is to measure how similar a cell is to its surrounding cells in order to determine if they can be joined together in a neighborhood.
To do so, we devise a homogeneity measure defined here as \texttt{H\_Index}, which we calculate for each cell. 
For each cell $p \in Grid$, we collect all the cells within a certain radius $r$ of that cell, creating a set of neighboring cells $N_{p,r}$. Then, we calculate the \texttt{H\_Index} value with respect to $p$ and $r$, which we formally define as:
\begin{displaymath}
\mathrm{H\_Index(p,r)} = \frac{\sum_{n \in N_{p,r}} \mathrm{cos(\mathbf{v_p},\mathbf{v_n})} \times \mathrm{smooth(n)}} {|N_{p,r}|}
\end{displaymath}
where $\mathrm{cos}(\mathbf{v_p}, \mathbf{v_n})$ is the cosine similarity between the vectors of the two grid cells $p$ and $n \in N_{p,r}$, and $\mathrm{smooth(n)}$ is a 2-dimensional isotropic Gaussian smoothing function of size $r^2$ centered on $p$. We average the weighted similarity over all the grid cells $n \in N_{p,r}$. 
We can choose various radii $r$ around each cell to compute \texttt{H\_Index} in order to find homogeneous areas at different scales. Figure~\ref{fig:Rad} shows the heatmaps for New York when we plot the \texttt{H\_Index} value with a radius of $400$, $800$, and $1200$ meters, respectively. 
At $400$ meters, we see many small pockets of areas with relatively high \texttt{H\_Index} values but as the radius increases, the heatmap becomes dominated by geographically broader groups. This demonstrates how we can use the radius measure to find areas of relatively high homogeneity at varying scales. During our evaluation in Section~\ref{sec:evaluation}, we present the neighborhoods we have detected using radii of $400$, guided by what has been established as a standard in the urban planning research community~\cite{mehaffy2010}. However, in the recommendation scenario presented in Section~\ref{sec:application} we experiment with radii of $400$ and $800$ meters as we explore trade-offs that are relevant to the application under consideration.   

We experiment with other similarity measures to define \texttt{H\_Index}, including a simple binary cosine similarity (as opposed to using density values), which records $1$ for the presence of a cluster at the point and $0$ otherwise, a Jaccard similarity coefficient, and vector intersection. In Figure~\ref{fig:Sim}, we see how the heatmap which uses the cosine similarity measure with density compares with the other measures. We choose the density-based cosine similarity as it yields the best results experimentally, as the simple binary cosine similarity and Jaccard similarity coefficient are biased towards areas with less data, while vector intersection favors areas with the most data.
\subsection{Detecting Neighborhood Boundaries}
\label{modelStep3}
Finally, given a heatmap of \texttt{H\_Index} values for every point in the city, $p \in Grid$, we wish to highlight areas of high relative homogeneity, that is, detect boundaries for our neighborhoods. However, we cannot simply set a threshold on \texttt{H\_Index} to find neighborhoods, because locally homogeneous areas do not have the same maximum \texttt{H\_Index} values. Thus, we employ a moving threshold and a range of acceptable neighborhood sizes, \texttt{min} and \texttt{max}, for each radius $r$. In this way, we can also find neighborhoods at different sizes for each scale. Algorithm~\ref{bound} describes how we find neighborhoods for each heatmap.
\begin{algorithm*}
\caption{FindBoundaries(P, min, max)}\label{bound}
 \textbf{Input:} $P$: all $(lat,lng)$ cells $p \in Grid$\\
 $min$, $max$: acceptable range of neighborhood sizes
\\\textbf{Output:} $N$: list of groupings of cells in $Grid$ that have relatively high local homogeneity
\algrenewcommand\algorithmicfor{\textbf{for each}}
\begin{algorithmic}[1]
\While{$len(P) > 0$}
	\For{$p$ in $P$}
		\If{$p.H\_Index > threshold$}
			\State $PassedP.append(p)$
			\EndIf
	\EndFor
	\For{$p$ in $PassedP$}
		\If{$p.visited = false$}
			\State $GroupP[groupnum].append(p)$
			\State $VisitDFS(p, GroupP)$
		\EndIf
	\EndFor
	\For{$g$ in $GroupP$}
		\If{$g.length > min$ and $g.length < max$}
			\State $N.append(g)$
			\State $P.delete(g)$
		\EndIf
	\EndFor
	\State $threshold = threshold + increment$
\EndWhile \\
\Return $N$
\end{algorithmic}
\end{algorithm*}

Given all the (x,y) latitude and longitude tuples in the grid and an acceptable range of number of grid cells, \texttt{min} and \texttt{max}, required to form a neighborhood, 
we set the threshold value to $0$ and initialize an empty list \texttt{N} that will store our neighborhoods. We begin with the input \texttt{P}, which is the whole geographic area of the city represented as the set of grid cells.
In Step $2$, we iterate through the values in \texttt{P}, collecting all the cells that have \texttt{H\_Index} $>$ \texttt{threshold} and place them in the list \texttt{PassedP}. Initially, this will be the entire set \texttt{P} as the threshold is $0$. Then we find all geographically grouped cells in \texttt{PassedP} and add each group to the list \texttt{GroupP}. Two points are in the same group if there is a path from one to the other in \texttt{PassedP} by hopping 100 meters north, south, west, or east. Thus we can find groupings using a simple depth-first-search graph traversal algorithm. 
Once we have visited a node, we flag it as visited and continue until all cells in \texttt{PassedP} are visited.
In Step 13, we iterate through the groups of cells in \texttt{GroupP}. If the number of cells in the group is within the acceptable range, then we add the group to \texttt{N} and delete the cells in the group from \texttt{P}. Finally we increment \texttt{threshold} by $0.02$ and return to Step 2 with the remaining cells in \texttt{P}. We continue this until there are no more cells in \texttt{P}, at which point, we return \texttt{N}, a list of groupings of cells that have controlled neighborhood sizes and high local homogeneity. In Section~\ref{sec:application} we will demonstrate that the size of neighborhoods matters for certain applications. For instance, if we would like to use neighborhood polygons in the context of a recommendation task, then being able to explicitly control for area sizes would be a desirable property to suggest less or more geographically precise areas. In our experiments we have set the parameters $min$ and $max$ to be $8$ and $160$ respectively to allow for the detection of a diverse set of possible neighborhood sizes. 

\section{Evaluation}
\label{sec:evaluation}
\begin{figure*}
\centering
\includegraphics[scale=0.53]{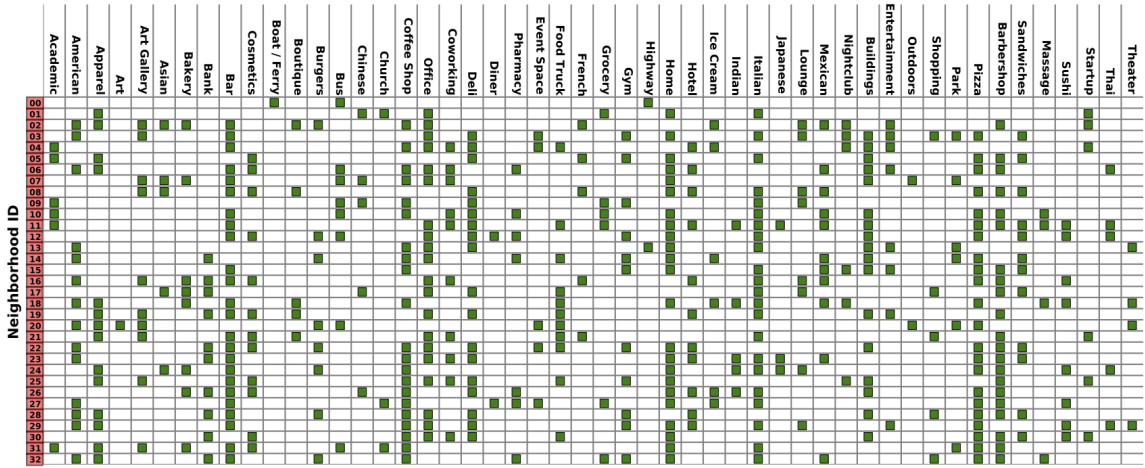}
\caption{Top-$20$ \emph{Place Type} Features for each neighborhood in New York City. Only a subset of the $50$ most popular place types in the city is shown due to space constraints.}
\label{fig:canalysis}
\end{figure*}
Having devised a set of techniques to extract urban activity features and neighborhood boundaries in cities we now evaluate them by analyzing the resulting neighborhoods in terms of their respective features. We also present an online tool named Hoodsquare, which draws the boundaries of hotspots and neighborhoods on an interactive map.
\subsection{Analyzing Neighborhood Features}
\begin{figure*}
\centering
\resizebox{6cm}{!}{\includegraphics{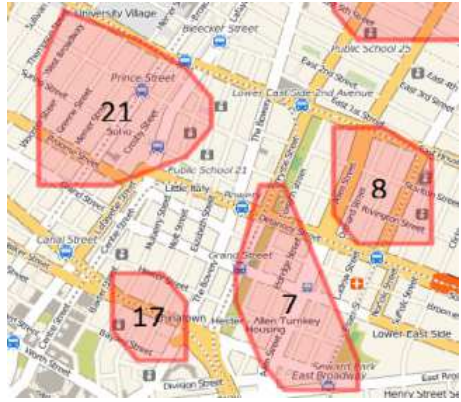}}
\caption{Neighborhoods near SoHo in New York.}
\label{fig:LES}
\end{figure*}

We now perform a feature-based analysis of the neighborhoods in New York found by our model.
Focusing on the \emph{Place Type} feature, in Figure~\ref{fig:canalysis} we depict the top-$20$ most prominent venue categories observed in each of the $32$ neighborhoods in the city. Some place types (e.g., Bar and Corporate Office) tend to be more present than others (e.g., Boat/Ferry) across neighborhoods. Also, a diverse set of neighborhoods has emerged after the application of the Hoodsquare algorithm, with certain areas having only a small set of features (see neighborhood  $00$), whereas most are geographic hotspots for numerous urban activities simultaneously, as noted by the presence of various types of venues within their boundaries. 
On a segment of New York near SoHo shown in Figure~\ref{fig:LES}, we can see the top 4 \emph{Place Type} features found in each neighborhood in Table 2. 
As it can be noted, our neighborhoods may represent very diverse areas that are at the same time geographically close to each other. In accordance to popular conception and observation, the area encompassed by neighborhood 21, commonly identified as SoHo, has many shops, while to the south, neighborhood 17, around Chinatown, has predominantly Chinese restaurants. Likewise, neighborhood 7, near the area of Bowery, has mostly art galleries, while neighborhood 8, which is closest to the Lower East Side, is mostly bars. There is also some overlap, with a strong Chinese restaurant presence in neighborhood 7. We also see a sizable portion of Italian restaurants in neighborhood 17, and yet no separate cluster for Little Italy, demonstrating the documented encroachment of Chinatown and SoHo on an ever-shrinking Italian neighborhood~\cite{littleitaly}. 
\subsection{Hoodsquare: A Map-based Tool for Urban Exploration}
\begin{figure*}[t]	
\centering
\includegraphics[width=1.35\columnwidth] {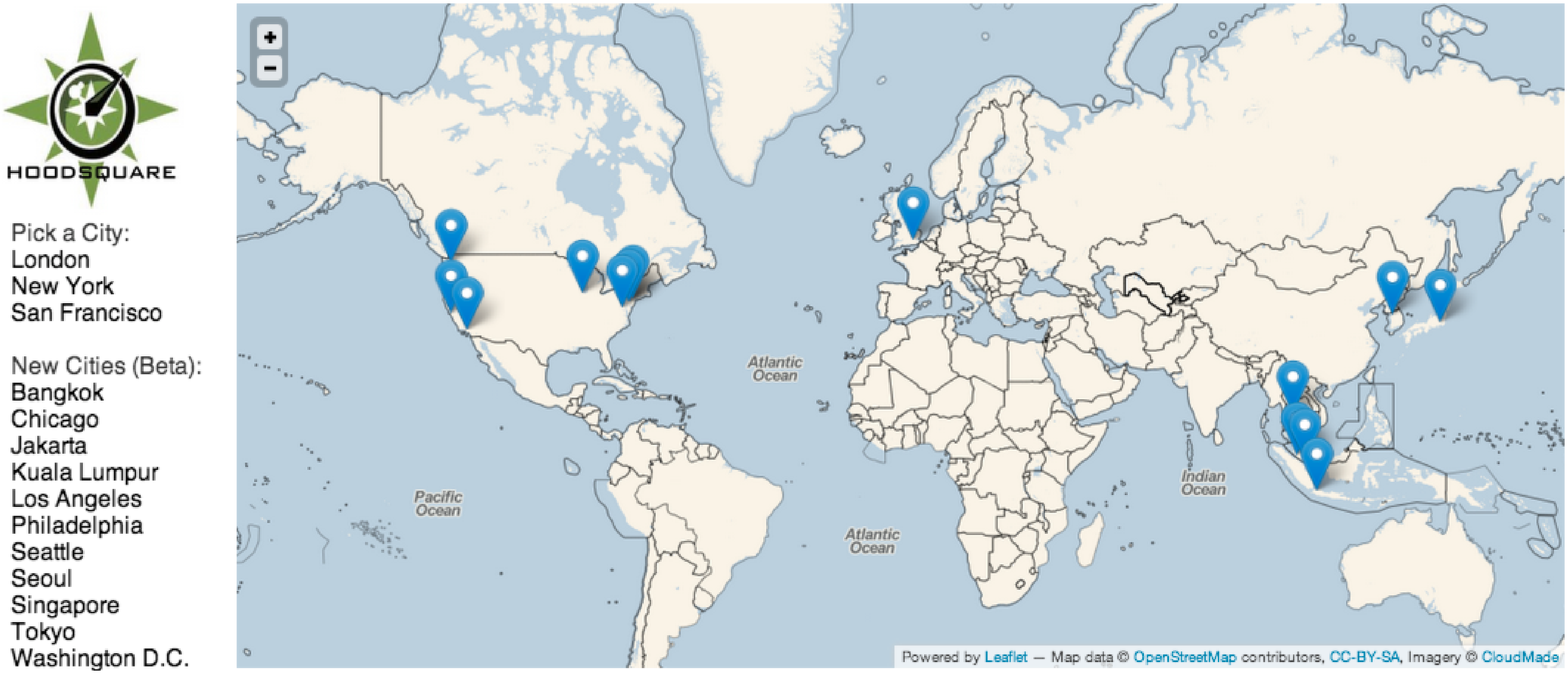}
\caption{The Hoodsquare web site menu: user can select the city they wish to explore through the map-based interface.}
\label{fig:menu}
\end{figure*}

\begin{figure*}[t]	
\centering
\includegraphics[width=1.0\columnwidth] {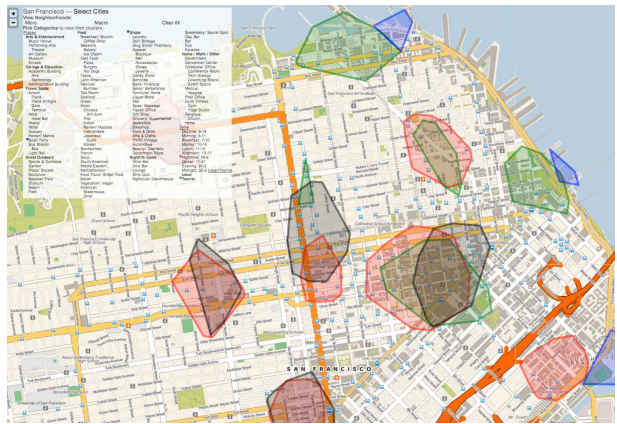}
\caption{Polygons of selected geographic activity hotspots in the city of San Francisco: users can click on the menu on the top left of the screen and
select a type of urban activity or to view neighborhoods.}
\label{fig:hotspots}
\end{figure*}

Guided by demonstration of the importance of urban recognizability~\cite{psychomaps}, we are motivated in our work to characterize and recommend geographic areas greater than single venues as a means for people to explore their local surroundings and independently discover areas of interest. With this sentiment, we have plotted both the feature hotspots and neighborhoods resulting from our methodology in a web vizualization named Hoodsquare. 
\begin{table*}
\centering
\label{tab:LTable}
{\small
\begin{tabular}{|l|l|l|l|l|}
\hline
ID&1\textsuperscript{st} Place&2\textsuperscript{nd} Place&3\textsuperscript{rd} Place&4\textsuperscript{th} Place\\
\hline
21&Apparel 12.6\%&Office 8.9\%&Boutique 5.9\%&Gallery 5.9\%\\
\hline
17&Chinese 27.7\%&Italian 7.9\%&Tea Room 5.0\%&Vietnamese 4.0\%\\
\hline
7&Gallery 11.5\%&Chinese 5.5\%&Park 3.6\%&Home 3.6\%\\
\hline
8&Bar 8.2\%&Home 5.1\% & Boutique 5.1\%&Rock Club 4.8\%\\
\hline
\end{tabular}}
\caption{Top 4 Place Type features for 4 different neighborhoods near the Lower East Side of New York.}
\end{table*}
First, users choose choose a city in the interface shown in Figure~\ref{fig:menu}.
The three cities discussed in Section~\ref{sec:analysis} have the full set of features while $11$ additional cities are in beta stage with only the hotspot feature enabled. 
After a user has picked the city he or she would like to explore, Hoodsquare centers and zooms the map on it. 
As seen in Figure~\ref{fig:hotspots}, a menu is deployed at the top left of the map and contains a detailed list of all urban activity features (PlaceType, Time 
and Local/Tourist) together with the option to display neighborhoods of different sizes. 
As they click on the menu, users are presented with colored polygons encapsulating the geographic areas that are hotspots for a given feature.  
The example in Figure~\ref{fig:hotspots} shows the hotspot polygons for \textit{PlaceType} features Boat/Ferry (blue color) and Shops (black), the \textit{Time} feature Nighttime (red) and the Tourist (green)
feature from the \textit{Local/Tourist} class. The specific urban activity features available to explore in each city may differ from city to city due to cultural, geographic, or organizational idiosyncrasies, 
as pointed out by our exploration of features in Section~\ref{sec:analysis}.
In the next section, we present a practical application scenario and demonstrate how the output of tools such as Hoodsquare can be used.  
Finally, the hotspot and neighborhood polygons are also available for download to the research community\footnote{http://www.hoodsquare.org/data.zip}. 

\section{Recommending Neighborhoods} 
\label{sec:application}
\begin{figure*}
\centering
{\includegraphics[width=1.3\columnwidth]{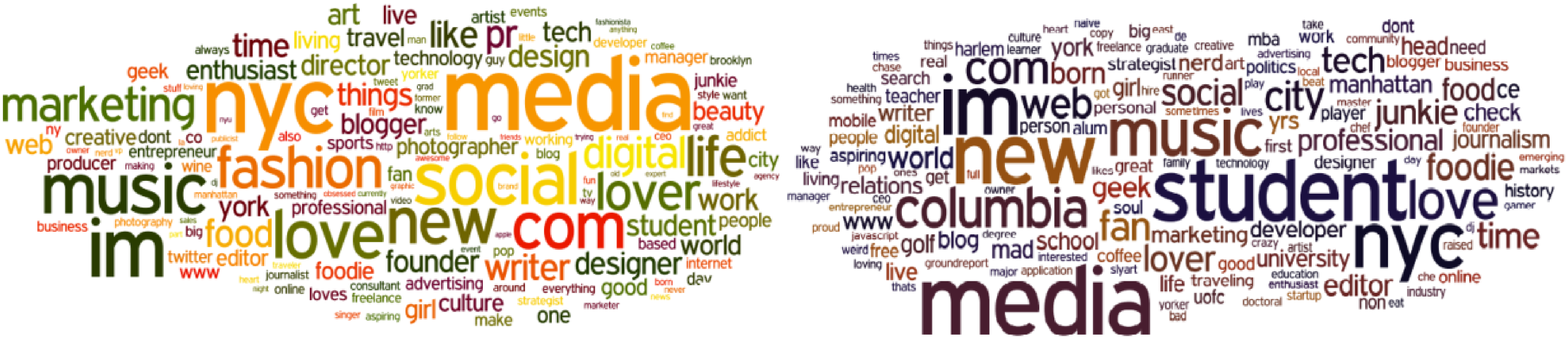}}
\caption{Word clouds of Twitter profiles of users that checked in to Neighborhood 21 (SoHo) and 9 (Columbia University).}
\label{fig:Soho}
\end{figure*}

In this section, we assess the neighborhood detection algorithm of Hoodsquare in a novel application scenario. 
Given a Foursquare user and textual information representing his or her public profile on Twitter, our goal is to determine the most relevant neighborhood in the city that the user may wish to visit (referred to here as the user's home neighborhood). However, besides recommending neighborhoods that match user profiles, our aim is also to achieve accurate recommendations with high geographic precision. That is, we would like to recommend neighborhoods that are small in terms of area size, so that users are guaranteed to be able to navigate in them with minimal physical and cognitive costs (when for instance they are searching for a restaurant). 
\newline
\\
\textbf{User Profiles Mapped to Neighborhoods.}
To measure the similarity between a user and a neighborhood based on textual data, we first describe a technique that allows us to represent neighborhoods as documents, based on the Twitter profiles of the users that check in there frequently. Focusing our evaluation on the city of New York, we can characterize neighborhoods by the types of people that visit them. The geographic congregation of individuals with similar personality and demographic traits has been theorized previously in~\cite{rentfrow2008theory}. 
We characterize a neighborhood by collecting all the users that have checked in there and aggregate the textual information available in their Twitter profiles. In Figure~\ref{fig:Soho}, we illustrate with a word cloud the terms we found in Neighborhood 21, around SoHo, and Neighborhood 9, near Columbia University. Using this information, we can gather an idea of the type of people that visit Neighborhood 21 in terms of their occupations, such as designer, PR, or marketing, and their interests, such as fashion, art, and food, whereas in Neighborhood 9, both occupations and interests reflect the academic nature of visitors or residents of the area. 
\newline
\\
\textbf{Training and Testing.} The evaluation of our recommendation scenario evaluation is defined in two separate stages. First, we use a subset of user profiles and define a training set in order to build our neighborhood documents. The remaining users form the test set and are the target users for recommendation. To achieve this, we gather all of the users from the city of New York and collect their Twitter profile text. Then we restrict the list of users to users with at least $5$ check-ins in the city and profile texts of at least $4$ terms, leaving us with $7511$ users (from a total $17,833$, see Table~\ref{tabledata}). We use $10$-fold cross validation to split our user list $10$ times into training and testing set sizes of $6760$ and $751$, respectively.
\newline
\\
\textbf{Predicting a User's Home Neighborhood.}
For the users in the testing set we aim to recommend the most relevant neighborhoods to them based on the textual data describing their profile.
In every cross-validation fold given a set of neighborhoods $\Gamma$, a test user $i$, and his home neighborhood $h_i \in \Gamma$ being the neighborhood where he has the most check-ins, our prediction model computes a similarity value $r_{i,j}$ for each candidate neighborhood $j \in \Gamma$. To compute the similarity score we use the cosine similarity measure between the user and neighborhood terms.
\newline
\\
\textbf{Evaluation Metrics.}  Having calculated the text-based vectors for each neighborhood and user we recommend to a test user the top-$N$ most similar neighborhoods and experiment with different recommendation list sizes $N$. We assess the different neighborhood boundary extraction strategies in light of two metrics. First, we assess by prediction accuracy, or Accuracy@$N$, which we refer to as the fraction of users whose home neighborhood was ranked at the top-$N$ of the recommendation list. Formally, for the test set of users $U_{test}$ and for the home neighborhood $h_i$ of user $i$ we define:
\begin{equation}
\text{Accuracy@}N = \frac{|P|}{|U_{test}|}, \text{where}~P=\{i: r_{i,h} \in \text{top-N}\}
\end{equation}
where $r_{i,h}$ is the cosine similarity score computed using user $i$'s profile information and his home neighborhood. Therefore, $P$ is the set of users for which the home neighborhood was ranked in the top-$N$ of the recommendation list. 

\begin{figure*}
\centering
\resizebox{8.0cm}{!}{\includegraphics{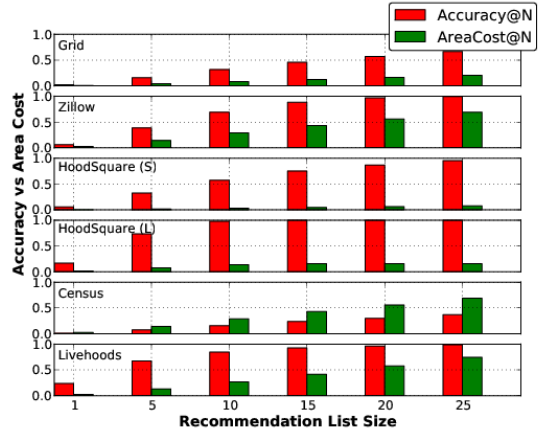}}
\caption{Comparison of the performance of different neighborhood sets using different list sizes. The $x$-axis shows to the prediction list size $N$ and the $y$-axis is prediction accuracy Accuracy@$N$ shown together with recommended area cost AreaCost@$N$.}
\label{fig:accuracy}
\end{figure*}

However, considering only prediction accuracy in this evaluation setting could lead to unfair comparisons. In the extreme scenario that the whole city
is recommended to a user as a single neighborhood, prediction accuracy would be always maximum ($1.0$). For that reason we employ another metric to account for the geographic precision of the recommendation, referred here as AreaCost@$N$ defined formally as:
\begin{equation}
\text{AreaCost@}N = \frac{\frac{\sum_{i}^{U_{test}} a_i}{|U_{test}|}}{A}
\end{equation}
where $a_i$ is the total recommended area, summing for the top-$N$ neighborhoods suggested to user $i$. Then we average the sum of all recommended areas across users, and we normalize with respect to geographic area of the city $A$ which in the case New York is $78km^{2}$ (Manhattan area). The scores yielded by AreaCost@$N$ range between $0.0$ and $1.0$ with values closer to $0.0$ referring to smaller neighborhoods and, thus, more precise recommendations. Finally, the goal for a recommendation algorithm will be to achieve a good trade-off between Accuracy@$N$ and AreaCost@$N$, when recommending neighborhoods to Twitter users.
\newline
\\
\textbf{Baselines.}
We consider several different boundary extraction methodologies in order to compare against the neighborhood boundaries we have extracted from Hoodsquare. The first is a set of neighborhood boundaries provided by Zillow, a web-based home and real estate marketplace~\cite{zillow}. A second baseline is the most recent census tract boundary data from $2010$ provided by the U.S. Census Bureau~\cite{census}. We also extract a simple grid baseline where we split the bounding box of the city into uniform squares of width $800$ meters. Finally, we include in our evaluation the neighborhood inference algorithm presented in the Livehoods project~\cite{Cranshaw_2012}, an approach that has considered a spectral clustering algorithm applied on the network of Foursquare places in the city. 
Each yielded cluster contains a set of places and the corresponding neighborhood is the bounding box extracted from the set of places that belong to a cluster.
Finally, with regards to Hoodsquare, we explore two different outputs: a set with small sized neighborhoods named \texttt{Hoodsquare S} and a set of larger neighborhoods \texttt{Hoodsquare L} that were produced by two different runs of the algorithm, one with a radius in \texttt{H\_Index} (see Section~\ref{sec:model}) of $400$ meters and another with $800$ meters. 
\newline
\\
\textbf{Results.}
In Figure~\ref{fig:accuracy} we present the results of the neighborhood recommendation task in terms of prediction accuracy and geographic precision for different values of list sizes $N$. In terms of prediction accuracy, the \texttt{Livehoods} and \texttt{Hoodsquare L} cases do the best by correctly ranking the home neighborhood with accuracy $23\%$ and $18\%$ of the users at the top position in the recommendation list (top-$1$). As we increase the list size $N$ to values $5,10,15,20$ and $25$, respectively, the results improve significantly for all strategies. Nonetheless, the increase of the size of the recommendation list implies a loss in geographic precision as the area cost of recommendations AreaCost@$N$ goes up. Notably the case of small neighborhoods (denoted as \texttt{Hoodsquare S}) manages to achieve relatively high accuracy (for instance $54\%$ for $N=10$) while attaining high geographic precision scores (only $2.41km^{2}$ for $N=10$ ). With regards to the \textit{naive} strategies (\texttt{Grid, Zillow, Census}) we note that the grid case does best, by managing a balanced trade-off between accuracy and area cost,
although Zillow achieves the highest accuracy scores as $N$ grows bigger. 
This behavior can be explained by the fact that in none of these cases the extraction of boundaries is optimized with respect to Foursquare user check-ins. On the contrary, \texttt{Zillow} and \texttt{Census} were derived with administrative purposes in mind. 

Taking a closer look at the comparison between \texttt{Hoodsquare} and \texttt{Livehoods} we have seen that the latter approach is lacking in terms of geographic precision. That is due to the large neighborhoods output by the algorithm. The \texttt{Livehoods} algorithm operates on a similarity matrix, mapping a network of places in the city. Two places which may be close in terms of network distance, could be far away geographically. This could result in very large neighborhood bounding boxes as the pairwise distance between venues in the neighborhood increases. This not only can have a negative effect in terms of precision in similar recommendation scenarios, but it could also violate the principle of homogeneity that defines the notion of the neighborhood as it has traditionally been established by the urban planning community. 
The \texttt{Hoodsquare} algorithm presented in Section~\ref{sec:model} alleviates these issues by building upon two main principles. The homogeneity index \texttt{H\_Index} similarity function is built specifically to capture the vibe of a geographic area, as this is reflected by the presence of Foursquare venues, local or touristic activity, and nearby temporal dynamics. Further, at the heart of the algorithm is a mechanism for geographic navigation, where points are added in the neighborhood by hopping $100$ meters east, west, north or south. Coupling this mechanism with the two parameters (\texttt{min} and \texttt{max}) that allow the experimenter to control for the geographic span of the detected neighborhoods, we are able to reach the desired output of small and homogeneous geographic areas. 

\section{Related work}
\label{sec:related}
A prominent source of neighborhood boundary data that has been used extensively in research is census tract or block data in the United States, collected by the U.S. Bureau of the Census~\cite{census}.
However, these boundaries have changed little since their inception in the early 20th century, an indication that they may not be a reliable source for boundaries of neighborhoods today. Still, much research on neighborhoods uses census tract data because it is easy to access and replicate \cite{Weiss_2007}.
An alternative way used by researchers to define neighborhoods is to allow residents to define where their own neighborhood's boundaries lie. However, studies have shown that a person's perception of his or her neighborhood varies greatly according to his or her characteristics, such as education level or income level, and the characteristics of a neighborhood, such as age and ethnic composition or socioeconomic status \cite{Sastry_2002}. Though these studies may be useful for research questions related to social cohesion or interaction, they may not be applicable in a broad context \cite{Roux_2001}. 

Despite the fact that location-based services such as Foursquare have become popular only recently, a series of works that specifically aim to characterize urban neighborhoods exploiting these data have appeared. 
Latent Topic Modeling has been used with location-based social networks to highlight regions in geographic space with similar co-occuring keywords \cite{Cranshaw_2010}. Both Latent Dirichlet Allocation and $k$-means clustering were also used to spot areas that corresponded to conceptual neighborhoods in San Francisco \cite{Chang_2011}. However, none of these studies attempted to delineate the boundaries of neighborhoods. 
Most recently, Cranshaw et  al.~\cite{Cranshaw_2012} in the Livehoods project we have already introduced in Section~\ref{sec:application}, segmented a city into areas according to Foursquare user check-ins. By applying a spectral clustering algorithm on the network of places in the city (two places are connected if they share a user), boundaries were generated by extracting the corresponding polygon for each cluster of places.
Our take in the task of detecting urban neighborhoods has two fundamental differences. First we do not just consider transitions between venues as sufficient to connect two areas; instead the similarity between geographic points factors in multiple signals (time, place types, tourist or local presence). Second, in the case of Livehoods, geographic clusters that represent neighborhoods are extracted by detecting components in a place network matrix, whereas Hoodsquare builds on a mechanism of geographic navigation through local hopping. This mechanism prevents it from yielding large neighborhoods that could be undesirable in a number of scenarios, including the mobile recommendation scenario we have examined. 
%
%
%

\section{Conclusion}
\label{sec:conclusion}
In this paper,  we have tackled the problem of generating neighborhood boundaries from location-based social media data. By collecting spatio-temporal venue and human mobility data from a check-in dataset collected from Foursquare, we characterize geographic spaces based on the types of venues nearby, the time of day they are busiest, and the presence of locals or tourists.
These feature-based representations form the input for a neighborhood detection algorithm. The algorithm exploits a homogeneity metric and a mechanism of geographic hopping to group similar areas together to yield neighborhoods. We have integrated the features and algorithms into a map-based tool named Hoodsquare.
In addition, we have showcased a user-to-neighborhood recommendation scenario, where neighborhoods are recommended to Twitter users based on textual profile information. We demonstrate that the neighborhoods we detect are not only relevant in terms of user interests, but also geographically coherent when compared to alternative techniques that construct neighborhoods using other means.
In terms of future work, we aim to explore new features that may be exploited to characterize urban spaces and also to expand the Hoodsquare tool to more cities and check-ins.

\section{Acknowledgements}
We acknowledge the support of EPSRC through grant GALE (EP/K019392) and Justin Cranshaw for the provision of the neighborhood
boundaries of the livehoods project.
\bibliographystyle{plain}
\bibliography{biblio}
\end{multicols*}
\end{document}